\documentclass[twocolumn,showpacs,preprintnumbers,
               superscriptaddress]{revtex4}
\usepackage{graphicx}

\renewcommand{\vec}[1]{\mbox{\boldmath $#1$}}
\newcommand{\at}{a_{\tau}}
\newcommand{\as}{a_{\sigma}}
\newcommand{\hf}{\mbox{-}}

\begin{document}

\preprint{HUPD-0120, YITP-02-19, RCNP-Th02004}
\preprint{hep-lat/0203025}

\title{Heavy quark action on the anisotropic lattice}

\author{Junpei~Harada}
\affiliation{Department of Physics, Hiroshima University,
             Higashi-hiroshima 739-8526, Japan}

\author{Hideo~Matsufuru}
\affiliation{Yukawa Institute for Theoretical Physics,
             Kyoto University, Kyoto 606-8502, Japan}

\author{Tetsuya~Onogi}
\affiliation{Yukawa Institute for Theoretical Physics,
             Kyoto University, Kyoto 606-8502, Japan}

\author{Ayumu~Sugita}
\affiliation{Research Center for Nuclear Physics, Osaka University,
             Ibaraki 567-0047, Japan}

\date{March 23, 2002}


\begin{abstract}
We investigate the $O(a)$ improved quark action on anisotropic
lattice as a potential framework for the heavy quark,
which may enable precision computation of hadronic matrix
elements of heavy-light mesons.
The relativity relations of heavy-light mesons as well as of
heavy quarkonium are examined on a quenched lattice with
spatial lattice cutoff $a_\sigma^{-1} \simeq$ 1.6 GeV and
the anisotropy $\xi=4$.
We find that the bare anisotropy parameter tuned for the massless quark
describes both the heavy-heavy and heavy-light mesons within 2\%
accuracy for the quark mass $a_\sigma m_Q  < 0.8$, which covers
the charm quark mass.
This bare anisotropy parameter also successfully describes
the heavy-light mesons in the quark mass region $a_\sigma m_Q \leq 1.2$
within the same accuracy.
Beyond this region, the discretization effects seem to grow gradually.
The anisotropic lattice is expected to extend by a factor $\xi$ 
the quark mass region in which the parameters in the action tuned for
the massless limit are applicable for heavy-light systems with well
controlled systematic errors.
\end{abstract}

\maketitle

\section{Introduction}
  \label{sec:Introduction}

Recent experimental progress in flavor physics to look for 
the effect of new physics requires precise theoretical 
predictions from the standard model.
However, model independent calculation of hadronic matrix elements
is difficult because of nonperturbative nature of QCD.
The lattice QCD simulation is one of most promising approaches
in which the systematic uncertainties can be reduced systematically
\cite{Lattice}.
The ultimate goal of this work is to construct a framework
for lattice calculations of hadronic matrix elements in a few percent
systematic accuracy, as required by the experiments in progress.
More practically, the level of accuracy requested from the CLEO-c
experiment \cite{CLEOC} as well as from the B Factories
\cite{KEKB,SLACB} are about 2 \%.
This paper intends to propose a systematic program to achieve
this accuracy making use of the anisotropic lattice QCD.

In a lattice calculation of heavy quark systems such as
the charm and bottom,
one needs to control the large discretization
error of $O((a m_Q)^n)$.
Extensive studies in various approaches have 
achieved much progress in understanding
the heavy quark systems.
However, it still seems difficult to achieve aforementioned systematic
accuracy with the current techniques for several reasons.
Here we categorize them into three types and summarize their
advantages and disadvantages from the viewpoint of
precision study of matrix elements in flavor physics.

{\it 1) Effective theories.}
One approach is to use descriptions of the heavy quark
based on the heavy quark effective theory using
nonrelativistic QCD \cite{TL91}.
As an advantage, the action incorporates higher order correction
in spatial momenta and can easily remove the all the mass dependent
errors at the tree or one-loop level.
However, since the theory does not have the continuum limit,
one cannot remove the discretization errors
by extrapolating the results on lattices with finite spacings.
Another disadvantage is that the nonperturbative 
renormalization is difficult due to
strong mass dependence.
Precise measurements of the weak matrix elements for the heavy-light
mesons below 10\% are therefore difficult.

{\it 2) Relativistic framework.}
The most straightforward approach is to use the $O(a)$ improved Wilson 
action in natural way for relatively lighter heavy quark,
and to extrapolate the results in $1/m$ according to
the heavy quark effective theory.
Since the theory has the continuum limit, the discretization errors
can be removed by an extrapolation. 
The perturbative errors can also be avoided by employing the
nonperturbative renormalization.
In practice, however, lattice artifacts of $O((a m_Q)^2)$ give dominant
errors \cite{KS02} so that the precise calculation will remain difficult
for the next few years even in quenched approximation.
Brute force improvement by decreasing lattice spacing and employing
large lattices quickly increase the simulation cost, and therefore
not a realistic solution for the calculation of matrix elements.

{\it 3) Fermilab approach.}
The Fermilab approach links the above two approaches
\cite{EKM97,Sro00}.
In the heavy quark mass region, the $O(a)$ improved Wilson action 
with asymmetric parameters is reinterpreted as
an effective theoretical description just like
the nonrelativistic QCD action.
Since the action reduces to the conventional $O(a)$ improved Wilson
action for small masses $a m_Q \ll 1$, it has in principle smooth 
continuum limit. 
The disadvantage is that it is not known how to extrapolate the
results on lattices with $a m_Q > 1 $, which are currently unavoidable
in particular for systems with $b$-quark.
For such a quark mass region, a mass dependent tuning of parameters
in the action is required for proper improvement, although
a systematic tuning beyond the perturbation theory is still
a challenge \cite{AKT01}.
Therefore precise calculation below 10\% level are nontrivial
for the weak matrix elements of the heavy-light mesons.

Therefore, it is desirable to develop a new framework for the heavy
quark with the following features.
(i) The continuum limit can be taken;
(ii) A systematic improvement program, such as the nonperturbative
 renormalization technique \cite{NRimp}, can be applied not only to
 the parameters in the action but also to the operators;
(iii) A modest size of computational cost is required for
 systematic computation of matrix elements.

The anisotropic lattice, on which the temporal lattice spacing
$a_{\tau}$ is finer than the spatial one $a_\sigma$,
is a candidate of such a framework \cite{Kla98a,Aniso01a}.
The approach is fundamentally along the line of Fermilab approach
on the anisotropic lattice.
However, large temporal lattice cutoff is expected to
improve the above problems drastically.
The most crucial one is that the mass dependence of the parameters
in the action may become so mild in the region of practical interest
that one can adopt the $O(a)$ improving clover coefficient determined
in the nonperturbative renormalization technique.
On the other hand, standard size of spatial lattice spacing
keeps a total computational cost modest.
Therefore, extrapolation of the result to the continuum limit
may be possible with keeping systematic uncertainties under
control with sufficient accuracy.
Whether these observations practically hold should be
examined numerically, as well as in the perturbation theory.

Our form of quark action on the anisotropic lattice 
is founded in Refs.~\cite{Aniso01a,Aniso01b}, 
where the perturbative results for light and heavy quarks 
and the simulation results of the light quark were studied.
In this paper, we focus on the heavy quark 
and study the mass dependence of the the Lorentz symmetry breaking 
effect in order to understand the mass range
for which a consistent description is possible.
This signals the applicability of anisotropic lattice to
the heavy quark systems.
We investigate the heavy-heavy and heavy-light mesons
on a quenched lattice of anisotropy $\xi=4$ and spatial
cutoff $a_\sigma^{-1}\simeq 1.6$ GeV.
The heavy quark mass is varied from the charm quark mass to
about 6 GeV to examine an applicability of the action to
these systems.

This paper is organized as follows.
In the next section, we first give our quark action,
which is discussed in detail in Ref.~\cite{Aniso01a}.
We then give our conjecture on the lattice spacing dependence of the 
anisotropy parameter toward the continuum limit, and discuss the
advantage of the anisotropic lattice compared to the isotropic one.
In Section~\ref{sec:expectation}, we observe the tree-level
expectation of the mass dependence of the $O(a^2)$ terms
in the quark dispersion relation and study how the anisotropic 
parameters or the breaking effect of relativity behave as a functions
of the heavy quark mass.
Section~\ref{sec:simulation} describes the numerical simulation.
We compute the heavy-light and heavy-heavy meson spectra and 
dispersion relations with two sets of heavy quark parameters.
In one set ({\it Set-I}), the bare anisotropy is set to the value
for the massless quark.
The other set ({\it Set-II}) adopt the result of the mass dependent
tuning using the heavy-heavy meson dispersion relation as obtained 
in Ref.~\cite{Aniso01b}.
We observe the mass dependence of the 
renormalized anisotropy in order to probe the breaking of relativity
for heavy quark. We also observe how the inconsistency among the
binding energies of heavy-heavy, heavy-light and light-light mesons
\cite{SAnomaly1,SAnomaly2} grows as a function of mass.
The last part of this section discusses the hyperfine splitting of the 
heavy-light meson.
In Section~\ref{sec:conclusion}, we summarize the result of
simulation, and draw our perspective on further development of
this framework, as our conclusion in this paper.

\section{Formulation}
 \label{sec:formulation}

\subsection{Quark action}

We adopt the following quark action
in the hopping parameter form \cite{Aniso01a,Ume01}:
\begin{equation}
  S_F  =  \sum_{x,y} \bar{\psi}(x) K(x,y) \psi(y),
\end{equation}
\begin{eqnarray}
 S_F  &=&  \delta_{x,y}
  - \kappa_{\tau} \left[ \ \ (1-\gamma_4)U_4(x)\delta_{x+\hat{4},y}
 \right.
 \nonumber \\
 & &  \hspace{3cm}
      + \left. (1+\gamma_4)U_4^{\dag}(x-\hat{4})\delta_{x-\hat{4},y} \right]
 \nonumber \\
 & & \hspace{-0.8cm}
    -  \kappa_{\sigma} \sum_{i}
         \left[ (r\!-\!\gamma_i) U_i(x) \delta_{x+\hat{i},y}
   +(r\!+\!\gamma_i)U_i^{\dag}(x\!-\!\hat{i})\delta_{x-\hat{i},y} \right]
 \nonumber \\
 & & \hspace{-0.8cm}
    -  \kappa_{\sigma} c_E
             \sum_{i} \sigma_{4i}F_{4i}(x)\delta_{x,y}
    - r \kappa_{\sigma} c_B
              \sum_{i>j} \sigma_{ij}F_{ij}(x)\delta_{x,y},
 \label{eq:action}
\end{eqnarray}
where $\kappa_{\sigma}$ and  $\kappa_{\tau}$ 
are the spatial and temporal hopping parameters, and are related
to the bare quark mass and bare anisotropy as given below.
The parameter $r$ is the spatial Wilson coefficient, and the
parameters $c_E$ and $c_B$ are the clover coefficients for 
the $O(a)$ improvement.
Although the explicit Lorentz symmetry is not manifest due to the
anisotropy in lattice units, 
it can be restored  in principle for physical observables in physical units 
at long distances up to errors of $O(a^2)$ by properly tuning 
 $\kappa_{\sigma}/\kappa_{\tau}$, $r$, $c_E$ and $c_B$
for a given $\kappa_{\sigma}$.
The action is constructed in accord with the Fermilab approach
\cite{EKM97} and hence applicable to an arbitrary quark mass,
although a mass dependent tuning of parameters is difficult
beyond the perturbation theory.
This may be circumvented by taking $a_\tau^{-1}\gg m_Q$, with which
the mass dependence of parameters are expected to be small so that
$O(a)$ Symanzik improvement program for the heavy quark can be 
applied. To study whether this expectation really holds or not is 
the main subject of this paper.

In present study, we vary only two parameters $\kappa_\sigma$
and $\kappa_\tau$ with fixed other parameters.
We set the Wilson parameter as $r=1/\xi$ and
the clover coefficients as the tadpole-improved tree-level
values, $c_E= 1/u_{\sigma} u_{\tau}^2$, and $c_B = 1/u_{\sigma}^3$.
The tadpole improvement \cite{LM93} is achieved
by rescaling the link variable as
$U_i(x) \rightarrow U_i(x)/u_{\sigma}$ and  $U_4(x) \rightarrow
U_4(x)/u_{\tau}$, with the mean-field values of the spatial 
and temporal link variables, $u_{\sigma}$ and $u_{\tau}$,
respectively.
Instead of $\kappa_\sigma$ and $\kappa_\tau$,
we introduce $\kappa$ and $\gamma_F(=1/\zeta)$ as
\begin{eqnarray}
\frac{1}{\kappa} &\equiv& \frac{1}{\kappa_{\sigma} u_\sigma}
     - 2(\gamma_F+3r-4)
    = 2(m_0 \gamma_Fa  +4) , \nonumber \\
\gamma_F &\equiv& 1/\zeta  \equiv
              \kappa_\tau u_\tau / \kappa_\sigma u_\sigma .
 \label{eq:kappa}
\end{eqnarray}
The former controls the bare quark mass and 
the latter corresponds to the bare anisotropy.

\subsection{Mass dependence of anisotropy parameter}

\begin{figure*}
\includegraphics[width=8.8cm]{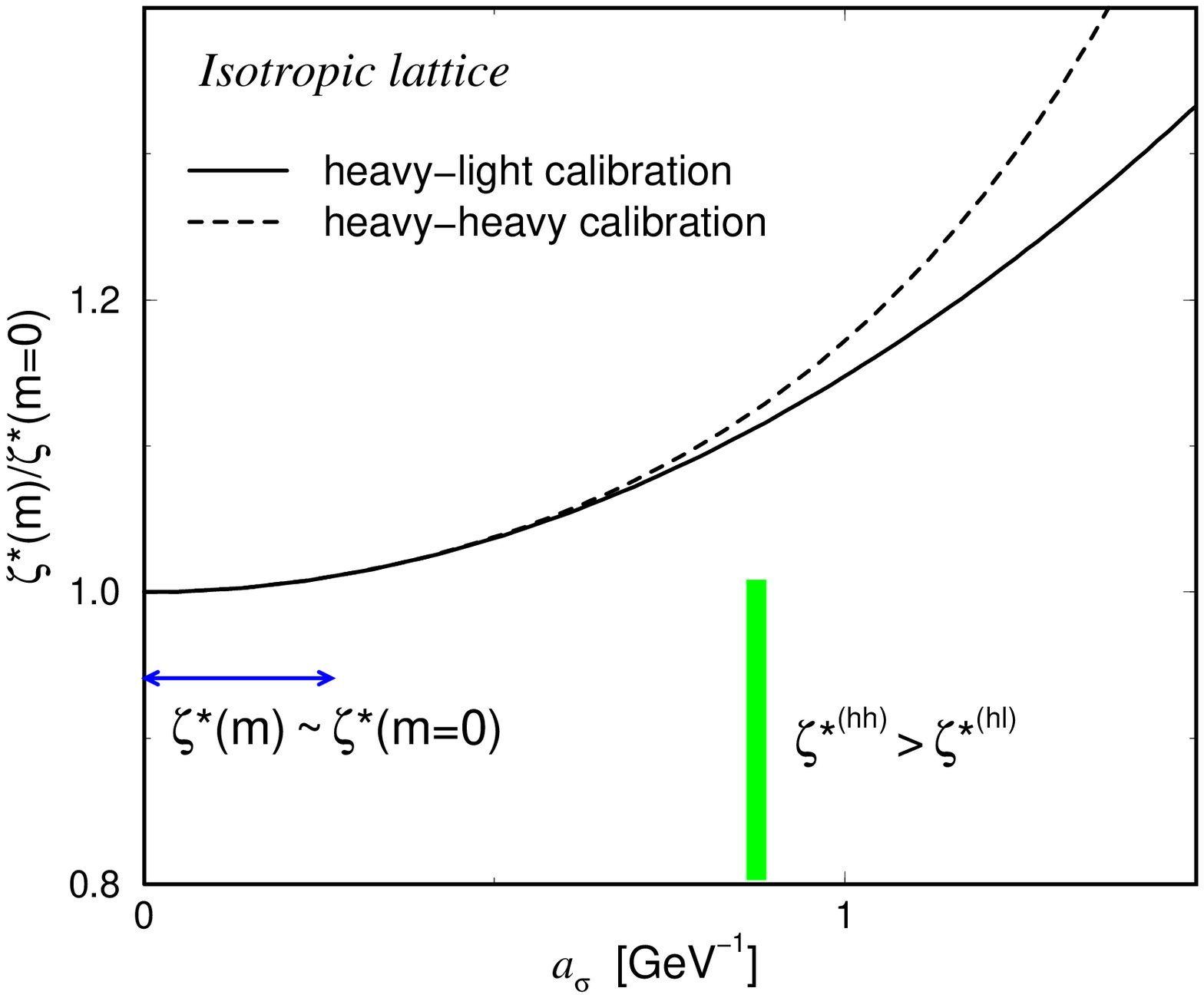}
\includegraphics[width=8.8cm]{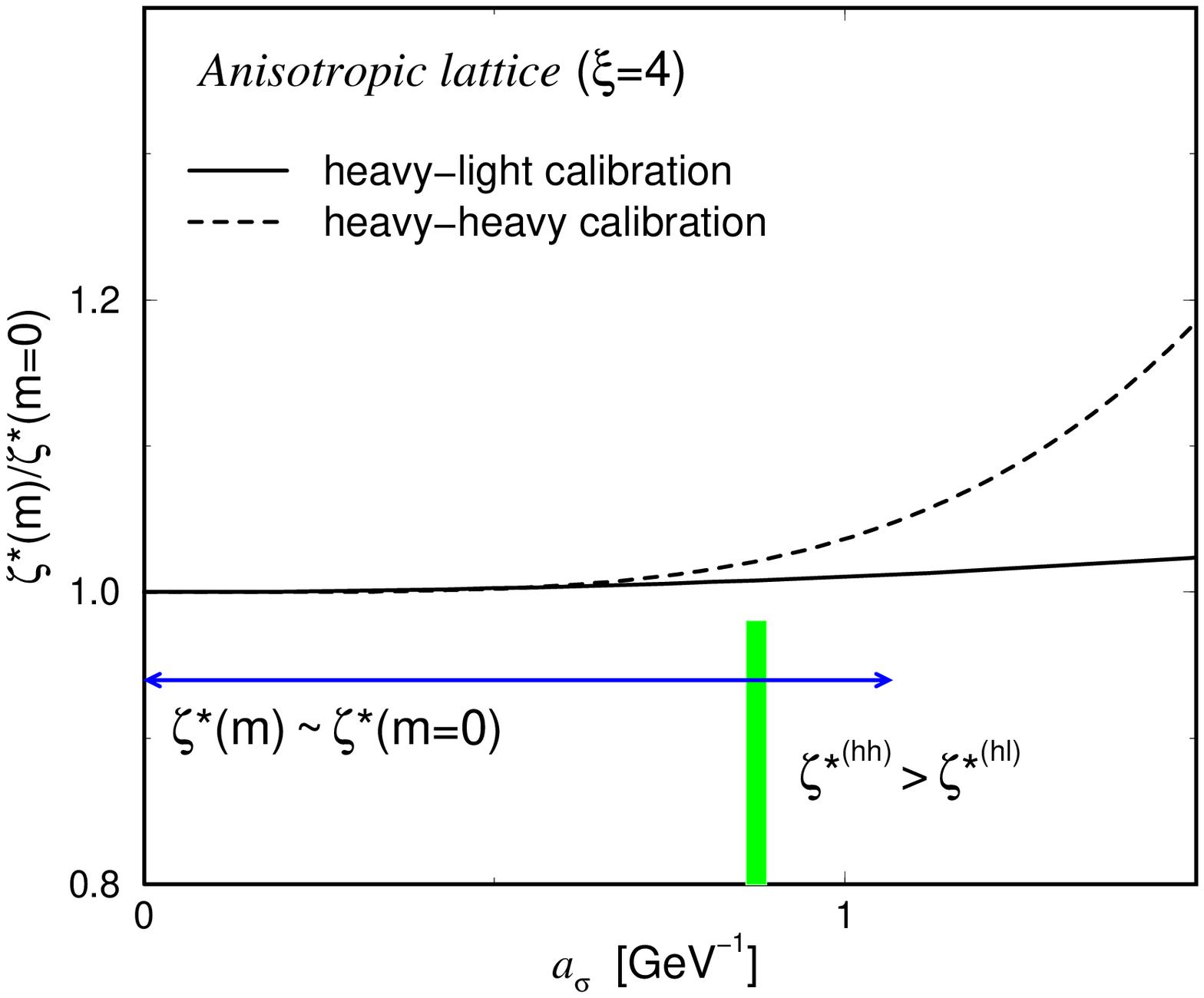}
\vspace{-0.6cm}
\caption{
The conjecture of lattice spacing dependence of the anisotropy
parameter for fixed quark mass, for isotropic (left)
and anisotropic (right) lattices.
Horizontal lines with arrows roughly represent the region
$a_\sigma < a_\sigma^{hl\hf ll}$, while vertical thick lines
correspond to $a_\sigma^{hh\hf hl}$, above which the $\zeta^*$'s
determined from heavy-heavy and heavy-light mesons no longer
equal. In this figure, a heavy quark with roughly the charm quark mass
is considered as an example.}
\label{fig:conjecture}
\end{figure*}

On an anisotropic lattice, one must tune the parameters so that 
the hadronic states satisfy the relativistic dispersion relations.
In general, the lattice dispersion relation for arbitrary values 
of $\zeta$ is described in lattice units as
\begin{equation}
E^2 (\vec{p}) = m^2 + \vec{p}^2/ \xi_F^2  + O(p^4).
\label{eq:DR1}
\end{equation}
In the above expression the energy $E$ and the rest mass $m$ 
are in temporal lattice units while the momentum $\vec{p}$ is
in spatial units.
The parameter $\xi_F$ introduced in this equation characterizes
the anisotropy of the quark fields.
The difference between the quark field anisotropy $\xi_F$ and
the gauge field anisotropy $\xi$ probes the breaking of relativity.
Therefore, the calibration is nothing but the tuning of
$\zeta(=1/\gamma_F)$ for a given $\kappa$ so that
$\xi_F$ equals $\xi$ and hence the relativistic dispersion 
relations are satisfied.
Let us call the tuned parameter as $\zeta^*(=1/\gamma_F^*)$.
$\zeta^*$ depends on the quark mass and can in general
be different from unity even for the case of isotropic lattice.
The requirement of the relativity relation automatically 
enforces that the rest and kinetic masses are equal to
each other.
In this sense, our calibration procedure of anisotropic lattice
action is a natural generalization of the Fermilab approach.

Now we consider the tuning of anisotropic parameter either on
isotropic or anisotropic lattice, for fixed physical quark mass.
The anisotropy parameter $\zeta=1/\gamma_F$
can be tuned using the dispersion relation of either heavy-heavy
or heavy-light meson in mass dependent way.
Alternatively, one can also adopt the value tuned for the
light-light mesons neglecting the mass dependence.
These procedures of calibration can give different results
due to the discretization errors.
We would like to know:
1) the value of lattice spacing $a_{\sigma}^{hh\hf hl}$
above which the calibrated parameters $\zeta^*$
using the heavy-heavy and heavy-light mesons differ from 
each other by more than a certain accuracy $\epsilon_{acc}$
(say, 2\%), and
2) the value of lattice spacing $a_{\sigma}^{hl\hf ll}$
above which the calibrated parameters $\zeta^*$ 
using the heavy-light and light-light mesons are different
by more than a certain accuracy $\epsilon'_{acc}$. 

Naively speaking, the difference between $\zeta^*$'s 
with heavy-heavy and heavy-light mesons originates from the
$O((a_{\sigma} p)^2)$ effects, where $p$ is the typical
quark momenta in heavy-heavy meson, $p \sim \alpha m_Q$.
On the other hand, the deviation of $\zeta^*$
with heavy-light mesons from the $\zeta^*$ with light-light
mesons is due to the
$O((a_{\tau}m_Q)^2)=O((a_{\sigma}m_Q/\xi)^2)$ effects.
With the criteria that these errors stay within the required
accuracies, we obtain
\begin{eqnarray}
a_{\sigma}^{hh\hf hl} & \sim & \sqrt{\epsilon_{acc}}/(\alpha m_Q) ,
  \nonumber \\
a_{\sigma}^{hl\hf ll} &  \sim & \sqrt{\epsilon'_{acc}} ~ \xi / m_Q .
\label{eq:a_crit}
\end{eqnarray}

Figure~\ref{fig:conjecture} conjectures the expected behavior
of the tuned anisotropy parameter $\zeta^*$ for
the isotropic ($\xi=1$) and anisotropic ($\xi=4$) lattices.
The tuned anisotropy normalized by the value at massless limit is
displayed for fixed physical quark mass, around charm quark mass, 
as an example.
In the case of isotropic lattice, as shown in the left panel of
Fig.~\ref{fig:conjecture},
$a_{\sigma}^{hl\hf ll} < a_{\sigma}^{hh\hf hl}$ is expected to
hold.
In this case, for a sufficiently small lattice spacing
$\as < a_{\sigma}^{hl\hf ll}$, both the heavy-heavy and heavy-light
mesons can be successfully described by setting the anisotropy
parameter to the value in the massless limit.
For a coarser lattice spacing
$a_{\sigma}^{hl\hf ll}< \as <a_{\sigma}^{hh\hf hl}$ ,
mass dependent tuning, namely the Fermilab approach, is necessary,
but the description of both the heavy-heavy and heavy-light mesons 
still works.
For an even coarser lattice spacing $a_{\sigma}^{hh\hf hl}<a_\sigma$,
simultaneously consistent description of heavy-heavy and
heavy-light mesons no longer works, because of severe discretization
effect in the heavy quarkonia.

Now let us consider the anisotropic lattice case with $\xi=4$.
The expected mass dependence of $\zeta^*$ is schematically represented
in the right panel of Fig.~\ref{fig:conjecture}.
As is obvious in Eq.~(\ref{eq:a_crit}) and will be studied in
the following sections, the anisotropy does not improve the situation
for heavy-heavy meson.
Therefore $a_{\sigma}^{hh\hf hl}$ remains roughly of the same size 
as in the isotropic lattice case.
On the other hand, Eq.~(\ref{eq:a_crit}) and 
the studies in the following sections show that for $\xi=4$,
$a_{\sigma}^{hl\hf ll}$ is expected to increase
and $a_{\sigma}^{hh\hf hl} < a_{\sigma}^{hl\hf ll}$ occasionally
holds.
This latter situation is particularly probable for quark with not
very large mass, and expected to hold up to the bottom quark mass
region.
In this case, for sufficiently small lattice spacing
$\as < a_{\sigma}^{hh\hf hl}$, both the heavy-heavy and heavy-light
mesons can be successfully described by setting the anisotropy
parameter to the value in the massless limit.
For a coarser lattice spacing
$a_{\sigma}^{hh\hf hl} < \as < a_{\sigma}^{hl\hf ll}$ consistent
description of heavy-heavy and heavy-light mesons no longer works.
Nevertheless the heavy-light meson is successfully described with
the anisotropy parameter tuned for the massless limit.
Finally for an even coarser lattice spacing $a_{\sigma}^{hl\hf ll}<\as$,
the genuine Fermilab approach, namely mass dependent tuning of $\zeta$,
is required to describe the heavy-light mesons.

In the range of lattice spacing $\as < a_{\sigma}^{hl\hf ll}$,
if we give up the description of heavy quarkonia and only consider 
the heavy-light systems, one can successfully adopt the tuned value
of $\zeta$ at the massless limit.
This is the most important region for our approach, since there
the mass dependence of parameters in the action other than
$\zeta=1/\gamma_F$, such as $c_E$ and $c_B$, are also expected
to be small, and hence the result of tuning at the massless limit
is also considered to be valid.
For the tuning of clover coefficient for the massless quark,
it may be possible to apply the nonperturbative renormalization
technique \cite{NRimp}, which is currently most efficient procedure
to remove the $O(a)$ effect beyond perturbation theory.
The most advantage of the anisotropic lattice is that this
region $a < a_{\sigma}^{hl\hf ll}$ is expected to be extended by
a factor $\xi$ as is found in Eq.~(\ref{eq:a_crit}),
so as to cover the heavy quark mass region of practical interests.
Therefore the anisotropic lattice is a candidate of
framework for heavy-light systems,
which satisfies the required characters (i), (ii), and (iii)
mentioned in the Introduction.
Also for heavy-heavy systems, it would be important to study
quantitatively the heavy quark mass at which the action starts
to fail in correct description.

This paper is dedicated to a  numerical study whether these
expectations actually holds.
Using Eq.~(\ref{eq:a_crit}), similar
expectation can be deduced for the heavy quark mass dependence 
with a fixed lattice spacing. 
In this paper, since we work with 
only one lattice spacing, instead of studying the lattice spacing 
dependence with a fixed heavy quark mass, we study the mass
dependence with a fixed lattice spacing. 
This means to specify in which region considered above the present
$a_\sigma$ is located for each value of quark mass.
This subject is discussed in more detail with the tree level analysis
in the next section, and with the numerical simulation
in the successive section.

\section{Expectation from tree level analysis}
\label{sec:expectation}

In this section, we review the tree level analysis of the heavy quark 
action in the Fermilab formulation.
In the following, we express the energy and the momenta in physical 
units so that the dependence on the lattice spacings $a_{\tau}$ and
$a_{\sigma}$ is explicitly shown.
The dispersion relation of the heavy quark on anisotropic lattice 
is given as \cite{EKM97,Aniso01a}
\begin{eqnarray}
\cosh( \at E)   &=&   1 +  \nonumber\\
& & \hspace{-2.2cm}
 \frac{[ \overline{m}_0 + 2 r \zeta \sum_i \sin^2(\as p_i/2)]^2
            + \zeta^2 \sum_i \sin^2(\as p_i) }
      {2 [ 1+ \overline{m}_0 + 2 r \zeta
                               \sum_i \sin^2(\as p_i/2)]},
\end{eqnarray}
where $\overline{m}_0=\at m_0$ .
This leads to the following dispersion relation.
\begin{eqnarray}
E^2 &=& M_1^2
   + \left(\frac{\xi^{\rm tree}}{\xi_F^{\rm tree}}\right)^2 \vec{p}^2
  \nonumber \\
  & & + A_1 \as^2 (\vec{p}^2)^2 + A_2 \as^2 \sum_i p_i^4 + \cdots,
\label{eq:Dispersion}
\end{eqnarray}
where $M_1$ is the pole mass,
\begin{eqnarray}
M_1 & =& \log( 1 + \overline{m}_0 ) \at^{-1},
\end{eqnarray}
and the anisotropy of the quark field at the tree level,
\begin{equation}
\left(\frac{\xi^{\rm tree}}{\xi_F^{\rm tree}}\right)^2
 = \log( 1 + \overline{m}_0 )  \xi^2  
 \left[ \frac{2 \zeta^2 }{\overline{m}_0 ( 2+\overline{m}_0)}
 + \frac{r \zeta}{1+\overline{m}_0} \right] .
 \label{eq:xi_F_xi}
\end{equation}
The third and fourth terms in Eq~(\ref{eq:Dispersion}) represent
the $O((ap)^2)$ errors, where $A_1$ and $A_2$ are given as
\begin{eqnarray}
A_1 & =& \frac{\xi^2}{4}
     \left[ \left( \frac{2\zeta^2}{\overline{m}_0 (2+\overline{m}_0)}
            + \frac{r\zeta}{1+\overline{m}_0} \right)^2 \right.
   \nonumber\\
&&          - \log(1+\overline{m}_0)              
         \left( \frac{8\zeta^4}{\overline{m}_0^3 (2+\overline{m}_0)^3}
       \right.
\nonumber\\
&& \left. \left.  
        + \frac{4\zeta^3[\zeta+2r(1+\overline{m}_0)]}  
            {\overline{m}_0^2 (2+\overline{m}_0)^2}
        +  \frac{r^2\zeta^2}{(1+\overline{m}_0)^2}\right) \right] ,
\end{eqnarray}
\begin{equation}
A_2  = -\frac{\xi^2}{3} \log(1+\overline{m}_0)
        \left[ \frac{2\zeta^2}{\overline{m}_0(2+\overline{m}_0)}
          +\frac{r\zeta}{4(1+\overline{m}_0)} \right].
\end{equation}
The above  coefficients are derived using 
the $M_1$, $M_2$, $M_4$ and $w_4$ in Ref.~\cite{EKM97} 
extended to anisotropic lattice by replacing certain
parameters as given in Ref.~\cite{Aniso01a}.

\begin{figure}
\includegraphics[width=8.8cm]{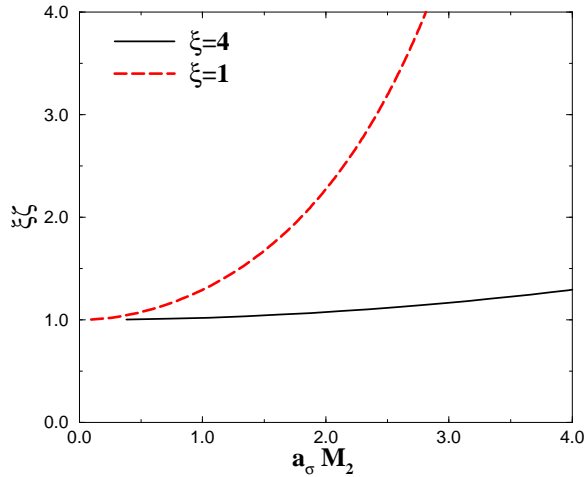}
\vspace{-0.8cm}
\caption{
The mass dependence of $\xi \zeta$ tuned at the tree level
for the isotropic lattice and anisotropic lattice with $\xi =4$.
The horizontal axis is the kinetic quark mass in spatial lattice
units.}
\label{fig:zeta}
\end{figure}

Fig.~\ref{fig:zeta} shows the mass dependence of $\zeta$ tuned
by requiring $\xi_F=\xi$, both on the isotropic lattice and
on the anisotropic lattice with $\xi=4$.
In the latter case,
it is clear that the mass dependence is drastically reduced so
that taking the value of $\zeta$ with massless limit is a
good approximation over a wide range of quark mass,
in contrast to the case of the isotropic lattice.

\begin{figure}
\includegraphics[width=8.8cm]{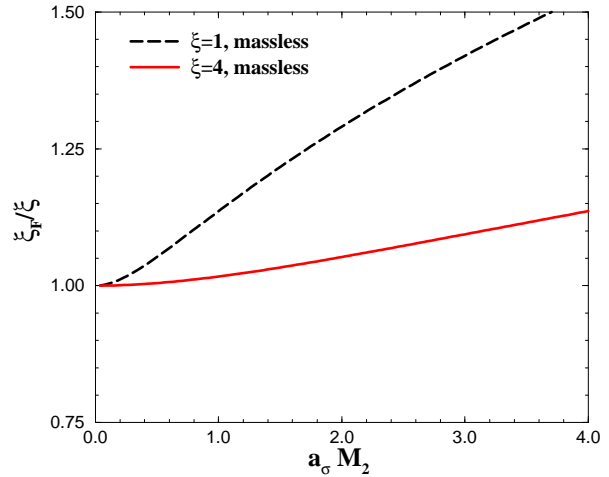}
\vspace{-0.8cm}
\caption{The kinetic mass dependence of $\xi_F/\xi$
at the tree level. We do not show the the result of mass dependent
tuning of $\zeta$ since $\xi_F/\xi=1$ is satisfied by definition.}
\label{fig:xi_F_xi}
\end{figure}

\begin{figure}
\includegraphics[width=8.8cm]{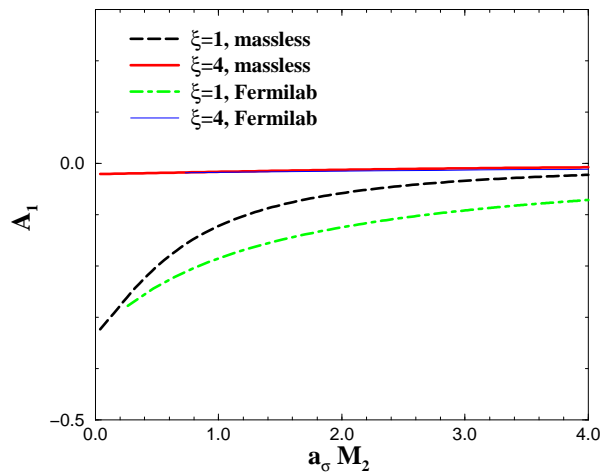}
\vspace{-0.8cm}
\caption{
The kinetic mass dependence of the $O((ap)^2)$ coefficient $A_1$
in the dispersion relation (\ref{eq:Dispersion}).
``massless '' denotes the result of tuning of $\zeta$ 
in the massless limit. ``Fermilab'' denotes the result of mass dependent
tuning of $\zeta$ by requiring $\xi_F=\xi$.}
\label{fig:A1}
\end{figure}

\begin{figure}
\includegraphics[width=8.8cm]{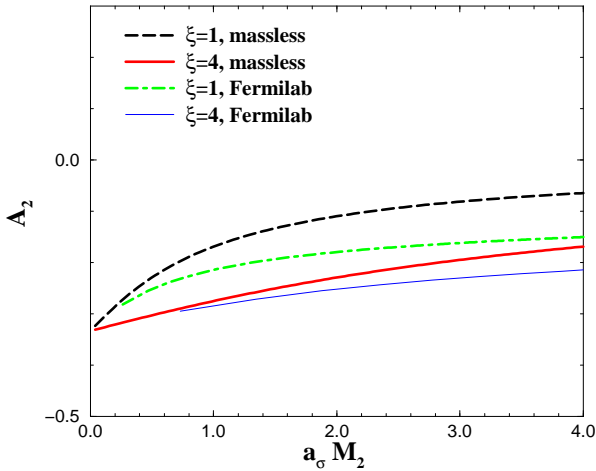}
\vspace{-0.8cm}
\caption{
The kinetic mass dependence of the  $O((ap)^2)$ coefficient $A_2$
in the dispersion relation (\ref{eq:Dispersion}).}
\label{fig:A2}
\end{figure}

In the following, let us consider either the case with mass
dependent tuning of $\zeta$ (denoted by ``Fermilab'' in the figures)
or with the $\zeta$ at the massless limit (``massless'').
Figure~\ref{fig:xi_F_xi} shows the breaking effect of relativity
relation, $\xi_F/\xi$, according to Eq.~(\ref{eq:xi_F_xi}).
The kinetic mass $M_2$ of quark is related to $\xi_F/\xi$
as $M_2=M_1 \xi_F^2/\xi^2$.
For the mass dependently tuned $\zeta$, $\xi_F$ trivially
equals $\xi$.
Without tuning, the mass dependence of $\xi_F$ is drastically
reduced in the case of $\xi=4$, as is expected from that
the deviation of $\xi_F/\xi$ from unity is the $O(\at m_Q)$ error.
In Figure~\ref{fig:A1}, one find that the coefficient $A_1$,
whose deviation from zero signals the $O(a^2)$ effect,
is also drastically reduced on the anisotropic lattice,
either with or without mass dependent tuning of $\zeta$.
On the other hand, 
$O(a^2)$ error from $A_2$ is not reduced but somewhat worse 
on the anisotropic lattice as is seen in Figure \ref{fig:A2}.
However, the coefficient $A_2$ is not larger 
than the value at the massless limit.  
As a general feature for all of the $\xi_F/\xi$, $A_1$ and
$A_2$, the mass dependences are drastically reduced 
on the anisotropic lattice.
On the other hand, mass dependent tuning of $\zeta$ on isotropic
lattice does not reduce the coefficients $A_1$ and $A_2$,
while it completely removes the discrepancy between $\xi_F$
and $\xi$.
In order to reduce these errors one has to introduce higher order
terms as is pointed out in Ref.~ \cite{EKM97}.

From these results, one can expect the following about the error 
of the heavy quark systems.
For the heavy-light systems, since heavy quark momenta typically
takes the size of the hadronic scale, $\Lambda_{QCD}\simeq$
200--500 MeV, the $O((ap)^2)$ errors 
are well under control even for large $\as M_2$,
if one keep $a_{\sigma}^{-1}\gg \Lambda_{QCD}$.
While the largest error originates from $A_2$, this is of the 
same order as that in the light quark systems.
For the heavy-heavy system, the situation is quite different.
Since the heavy quark momenta are typically $p\sim \alpha m_Q$,
The $O((ap)^2)$ errors are expected to be as large as
$A_1 (\alpha \as m_Q)^2$, $A_2 (\alpha \as m_Q)^2$.
Again the $A_2$ gives the largest contribution, but the size of 
the error is uncontrolled when $\as m_Q$ is large.

To summarize, the anisotropic lattice largely reduces the
discretization effects represented by $\xi_F/\xi-1$ and
$A_1$, while it does not improve the $A_2$.
For the heavy-light systems, this suffices for a computation
with discretization effects under control.
On the other hand, when $\as M_2$ is of order of unity,
the anisotropic lattice does not improve the situation for heavy
quarkonia, because of severe effect of $A_2$ in these systems.
Although mass dependent tuning of $\zeta$ further removes
the deviation of $\xi_F$ from $\xi$, the $\zeta$ tuned for
massless quark also stay a good approximation if
$a_\tau m_0 \ll 1$ holds.
This is the most advantage of anisotropic lattice compared with
the isotropic Fermilab approach, in which the mass dependence
of the parameter is much stronger.
These observations are in accord with our conjecture in the last
section for developing the anisotropic Fermilab formulation.

\section{Numerical simulation}
\label{sec:simulation}

This section numerically examines the idea described in
previous sections.
For this purpose, we perform simulations with two series of
heavy quark parameters.
For {\it Set-I} the anisotropy parameter is set to the value
at the chiral limit, $\gamma_F^*(m_q=0)$, just as same as for
the light quark.
The {\it Set-II} adopts the fully tuned anisotropy parameter
$\gamma_F^*$ using heavy-heavy mesons.
This calibration is done in the first subsection.
For the heavy-heavy and heavy-light mesons, the rest and kinetic
masses are obtained with two sets of parameters.
Two quantities are used to probe the breaking effect of the
relativity:
the fermionic anisotropy $\xi_F$ for heavy-heavy and heavy-light
mesons, and the inconsistency among the binding energies of
heavy-heavy, heavy-light, and light-light mesons.
These behaviors give us a hint to judge the regions in which our
framework can be applied.
The last subsection treats the heavy-light meson spectrum, and
observe how the hyperfine splitting behaves with meson kinetic mass.

\subsection{Calibration in heavy quark region}

The simulation parameters used in this paper is the second set
in Ref.~\cite{Aniso01b}:
the quenched anisotropic lattice of the size $16^3\times 128$
generated with the standard plaquette gauge action with
$(\beta,\gamma_G)=(5.95,3.1586)$, which correspond to the
renormalized anisotropy $\xi=4$ \cite{Kla98b}
and spatial lattice cutoff $a_\sigma^{-1}=1.623(9)$ GeV set by
hadronic radius $r_0$ \cite{Som94}.
The mean-field values in the quark action are set to
the mean values of link variables in the Landau gauge:
$u_\sigma=0.7917$ and $u_\tau=0.9891$.

In Ref.~\cite{Aniso01b}, the optimum bare anisotropy $\gamma_F^*$
is determined using the dispersion relation of mesons with
degenerate quark masses, and the resultant values of $\gamma_F^*$ 
are well represented by a linear function
\begin{equation}
 \frac{1}{\gamma_F^*} = \zeta_0 + \zeta_2 m_q^2,
 \hspace{0.7cm}
 m_q=\frac{1}{2\xi}\left( \frac{1}{\kappa}
        - \frac{1}{\kappa_c} \right) ,
\label{eq:fit_calib}
\end{equation}
where for the present lattice $\zeta_0 = 0.2490(8)$,
$\zeta_2=0.189(15)$, and $\kappa_c=0.12592(6)$.

\begin{table*}
\caption{
Quark parameters and the result of calibration.
The result for $\kappa=0.1100$, $0.1020$, and $0.0930$
are taken from Ref.~\cite{Aniso01b}.
The number of configurations is 200 for each value of the hopping
parameter, except for $\kappa=0.1100$ for which the simulation is
carried out with 300 configurations.}
\begin{ruledtabular}
\begin{tabular}{ccccccc}
 $\kappa$ & $m_0$ & input $\gamma_F$ &
 $\gamma_F^{*(PS)}$  & $\gamma_F^{*(V)}$ & $\gamma_F^*$ &
 $\delta \gamma_F^*$ \\
\hline
0.1100& 0.1437& 3.90, 4.00& 3.945(28)& 3.946(36)& 3.946(29)& -0.001(20)\\
0.1020& 0.2328& 3.90, 4.00& 3.848(26)& 3.847(34)& 3.847(28)&  0.001(15)\\
0.0930& 0.3514& 3.70, 3.80& 3.688(23)& 3.687(29)& 3.688(25)&  0.000(11)\\
0.0840& 0.4954& 3.39, 3.48& 3.471(20)& 3.462(23)& 3.467(21)& 0.0092(60)\\
0.0760& 0.6521& 3.04, 3.15& 3.205(20)& 3.191(23)& 3.199(21)& 0.0139(45)\\
0.0700& 0.7930& 2.85, 2.95& 2.946(23)& 2.932(23)& 2.939(23)& 0.0134(52)\\
0.0630& 0.9914& 2.50, 2,60& 2.584(22)& 2.561(24)& 2.573(23)& 0.0237(35)\\
\end{tabular}
\end{ruledtabular}
\label{tab:calib}
\end{table*}

\begin{figure}
\includegraphics[width=8.8cm]{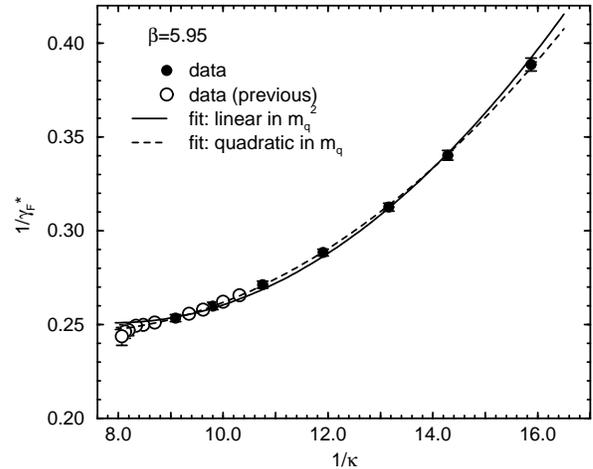}
\vspace{-0.8cm}
\caption{
The result of calibration in the heavy quark mass region
(filled circles), together with the previous result
in Ref.~\cite{Aniso01b}.
The fits are performed with all available data including previous
results.}
\label{fig:calib}
\end{figure}

In this paper, we use seven values of $\kappa$ for heavy
quark ($\kappa_h$), which roughly cover 1--6 GeV.
Three of them already appeared in Ref.~\cite{Aniso01b}.
We start with the calibration of remaining four values of
$\kappa$ in the heavy quark region in the same manner as
in Ref.~\cite{Aniso01b}.
The values of $\kappa$ used are listed in Table~\ref{tab:calib}
together with the result of calibration.
The second column is the naive estimate of bare quark mass
according to Eq.~(\ref{eq:fit_calib}).
For the heaviest case, $m_0$ is almost unity in temporal
lattice units, and therefore the breaking effect of relativity
may be visible.
Here we note that for heavier quark masses the difference
of $\gamma_F^*$ for pseudoscalar and vector mesons,
\begin{equation}
 \delta \gamma_F^* = \gamma_F^{*(V)} - \gamma_F^{*(PS)},
\end{equation}
is sizable beyond the statistical fluctuations.
This signals that the quarkonium system is not properly
described within the present framework at this lattice spacing.
This problem will again be discussed later in terms of the
fermionic anisotropies for heavy-heavy and heavy-light mesons.

Figure~\ref{fig:calib} shows the result of calibration.
The result is well fitted to a linear form in $m_q^2$
or a quadratic form in $m_q$, in spite of large quark mass.
The fits including previous data in Ref.~\cite{Aniso01b} result in
\begin{eqnarray}
\mbox{Linear:}    & &   \zeta_0 = 0.2510(6), \ \zeta_2 = 0.1437(26). \\
\mbox{Quadratic:} & & \zeta_0 = 0.2473(12), \ \zeta_1 = 0.0267(72),
 \nonumber \\
                  & &   \hspace{1cm}   \zeta_2 = 0.1151(81).
\end{eqnarray}
In both cases, the values of $\zeta_0$ is close to the value
at the mean-field tree level, $\xi^{-1}=0.25$.
Since there is no reason that the small quark mass dependence persist
up to $m_q \simeq a_\tau^{-1}$,
the difference between $\zeta_2$'s in linear fits with present data
and with only previous data in lighter quark mass region
is irrelevant, and above result just shows that the quark mass
dependence of $1/\gamma_F^*$ is still tractable at this mass region.

\subsection{Relativity breaking effect}

In this subsection we compute the heavy-heavy and heavy-light 
meson spectra and the dispersion relations 
using two sets of heavy quark parameters.
For the first one, {\it Set-I}, the bare anisotropy is set to
the value of $\gamma_F^*$ in the chiral limit obtained
in Ref.~\cite{Aniso01b},
namely $\gamma_F=4.016$ at $\beta=5.95$, for all quark masses.
The second one, {\it Set-II}, adopts the above result of mass
dependent calibration.
We use the same hopping parameter values $\kappa_h$ for the heavy quark 
as given in the previous subsection. 
For the light quark, we use single value, $\kappa_l =0.1235$.
The value of $\gamma_F$ at this $\kappa_l$ is set to the value in the
chiral limit, as same as in the light hadron spectroscopy
in Ref.~\cite{Aniso01b}.
We regard that at $\kappa=\kappa_l$ the quark mass is sufficiently light
for present purposes, and do not extrapolate to the chiral limit.
The numbers of configurations used are 200 and 500 for heavy-heavy and
heavy-light meson masses, respectively.

\begin{table*}
\caption{
Heavy-heavy meson spectrum for {\it Set-I} and {\it Set-II} parameters
obtained with 200 configurations.}
\begin{ruledtabular}
\begin{tabular}{ccccccccc}
 & $\kappa_h$ & $\gamma_F$ &
 $m_{PS}$ & $m_V$  &  $m_V - m_{PS}$  & 
 $\xi_F^{(PS)}$ &  $\xi_F^{(V)}$ &  $\xi_F^{(V)}-\xi_F^{(PS)}$ \\
\hline
{\it Set-I}
&0.1100& 4.016 & 0.42468(23)& 0.43755(33)& 0.01288(18)&
                                   4.069(34)& 4.068(46)& -0.001(24) \\
&0.1020& 4.016 & 0.58409(25)& 0.59358(34)& 0.00949(13)&
                                   4.148(27)& 4.143(35)& -0.005(16) \\
&0.0930& 4.016 & 0.76947(24)& 0.77692(31)& 0.00745(10)&
                                   4.292(22)& 4.294(28)&  0.002(10) \\
&0.0840& 4.016 & 0.96746(25)& 0.97358(31)& 0.00612(~8)&
                                   4.527(24)& 4.537(29)&  0.010(~8) \\
&0.0760& 4.016 & 1.15894(27)& 1.16411(33)& 0.00517(~8)&
                                   4.821(30)& 4.839(35)&  0.018(10) \\
&0.0700& 4.016 & 1.31453(27)& 1.31896(31)& 0.00443(~7)&
                                   5.146(33)& 5.187(38)&  0.041(~8) \\
&0.0630& 4.016 & 1.51262(27)& 1.51617(31)& 0.00355(~5)&
                                   5.650(39)& 5.726(43)&  0.076(~7) \\
\hline
{\it Set-II}
&0.1100& 3.946& 0.42942(23)& 0.44248(33)& 0.01306(18)&
                          4.005(29)& 4.003(41)& -0.002(22)\\
&0.1020& 3.847& 0.60172(25)& 0.61146(33)& 0.00975(14)&
                          4.002(24)& 4.000(32)& -0.003(14)\\
&0.0930& 3.688& 0.81797(24)& 0.82574(31)& 0.00777(10)&
                          3.995(19)& 4.000(24)&  0.005(~9)\\
&0.0840& 3.467& 1.07512(25)& 1.08167(30)& 0.00655(~7)&
                          3.996(18)& 4.005(22)&  0.009(~6)\\
&0.0760& 3.199& 1.35939(26)& 1.36509(31)& 0.00571(~7)&
                          4.002(22)& 4.011(26)&  0.010(~6)\\
&0.0700& 2.939& 1.62650(27)& 1.63164(31)& 0.00514(~6)&
                          3.993(23)& 4.009(26)&  0.016(~5)\\
&0.0630& 2.573& 2.02101(28)& 2.02550(31)& 0.00449(~5)&
                          3.988(23)& 4.013(26)&  0.025(~4) \\
\end{tabular}
\end{ruledtabular}
\label{tab:HHspc}
\end{table*}

\begin{table*}
\caption{
Heavy-light meson spectrum for {\it Set-I} and {\it Set-II} parameters
obtained with 500 configurations.}
\begin{ruledtabular}
\begin{tabular}{cccccccc}
 & $\kappa_h$ &  $m_{PS}$ & $m_V$  &  $m_V - m_{PS}$  & 
 $\xi_F^{(PS)}$ &  $\xi_F^{(V)}$ &  $\xi_F^{(V)}-\xi_F^{(PS)}$ \\
\hline
{\it Set-I}
&0.1100& 0.29108(27)& 0.30997(51)& 0.01889(39)&
                             3.985(36)& 4.006(49)& 0.020(44) \\
&0.1020& 0.37705(30)& 0.39106(50)& 0.01401(34)&
                             3.994(39)& 4.031(51)& 0.036(42) \\
&0.0930& 0.47595(36)& 0.48643(57)& 0.01048(36)&
                             4.032(51)& 4.101(73)& 0.070(55) \\
&0.0840& 0.58123(46)& 0.58123(46)& 0.00805(43)&
                             4.088(73)& 4.21(11) & 0.119(77) \\
&0.0760& 0.68279(55)& 0.68916(81)& 0.00638(44)&
                             4.148(99)& 4.30(15) & 0.147(87) \\
&0.0700& 0.76546(64)& 0.77079(89)& 0.00534(45)&
                             4.21(13) & 4.38(18) & 0.173(98) \\
&0.0630& 0.87080(78)& 0.8751(10) & 0.00430(47)&
                             4.30(18) & 4.51(25) & 0.21(12) \\
\hline
{\it Set-II}
&0.1100& 0.29323(27)& 0.31229(52)& 0.01906(39)&
                          3.952(36)& 3.970(48)& 0.018(44) \\
&0.1020& 0.38526(30)& 0.39961(50)& 0.01435(34)&
                          3.905(37)& 3.936(49)& 0.031(40) \\
&0.0930& 0.49889(35)& 0.50990(57)& 0.01101(37)&
                          3.846(46)& 3.904(66)& 0.058(51) \\
&0.0840& 0.63258(40)& 0.64135(60)& 0.00877(36)&
                          3.757(54)& 3.829(76)& 0.071(54) \\
&0.0760& 0.77918(49)& 0.78647(75)& 0.00729(43)&
                          3.666(73)& 3.77(11) & 0.108(71)\\
&0.0700& 0.91609(54)& 0.92246(80)& 0.00637(44)&
                          3.571(84)& 3.69(12)& 0.118(73) \\
&0.0630& 1.11749(62)& 1.12294(87)& 0.00544(45)&
                          3.45(10) & 3.58(14)& 0.130(77) \\
\end{tabular}
\end{ruledtabular}
\label{tab:HLspc}
\end{table*}

The meson spectra are listed in Table~\ref{tab:HHspc} and
\ref{tab:HLspc} for heavy-heavy and heavy-light mesons, respectively.
In these tables, we also list the results of $\xi_F$ for each
meson channel, and difference between them,
$\delta \xi_F = \xi_F^{(V)} - \xi_F^{(PS)}$.
If the anisotropic lattice action does not describe the quarks
inside mesons in respecting the relativity relation,
the breaking effect appears in the dispersion relations of
mesons.
Therefore the deviation of fermionic anisotropy $\xi_F$ from $\xi$ 
signals the breaking effect of relativity.
In the following, we first discuss on the result of $\xi_F$ for
{\it Set-I} parameters, namely with $\gamma_F$ tuned for massless limit,
and then briefly summarize the result for {\it Set-II} parameters.

\begin{figure*}
\includegraphics[width=8.8cm]{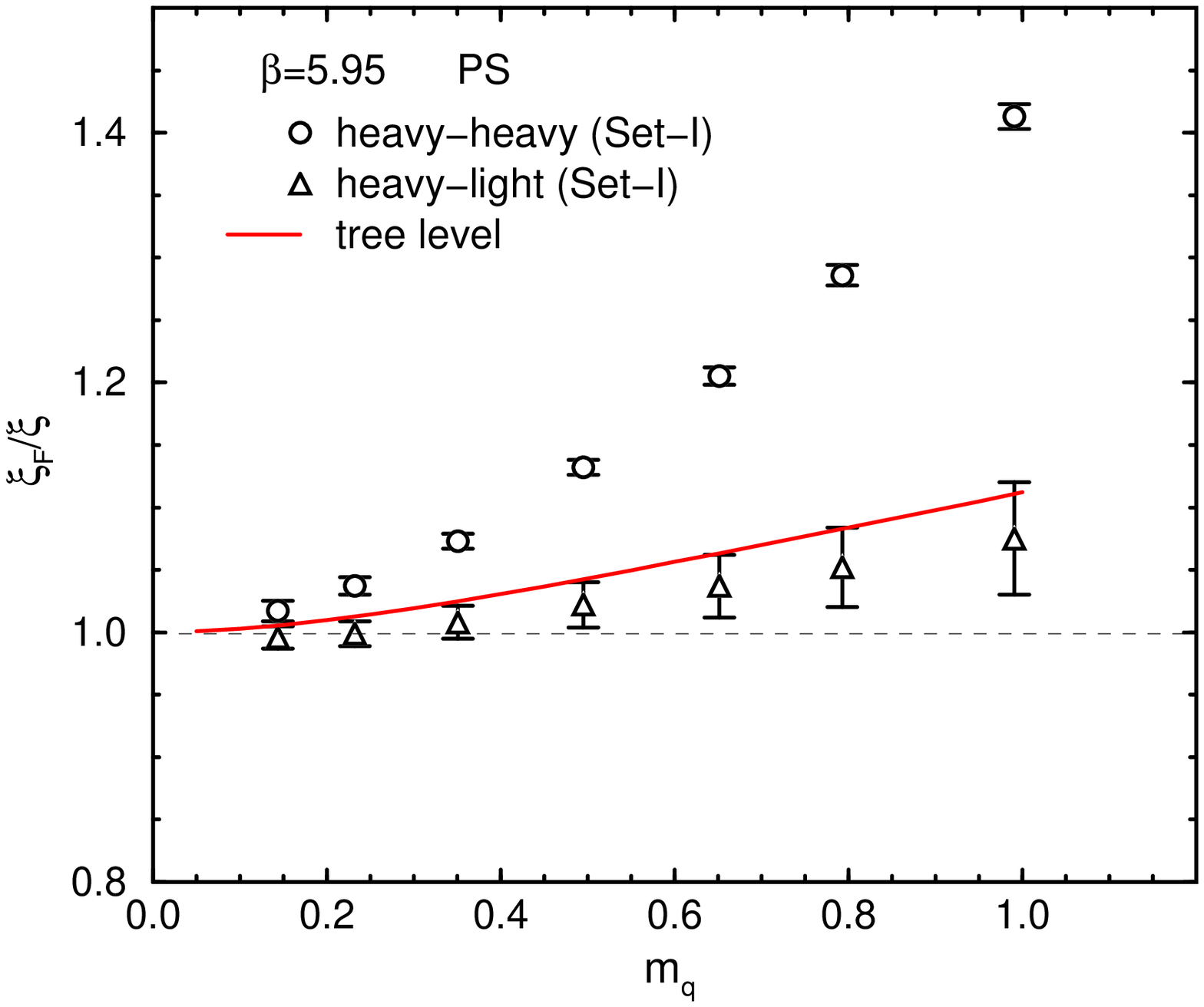}
\includegraphics[width=8.8cm]{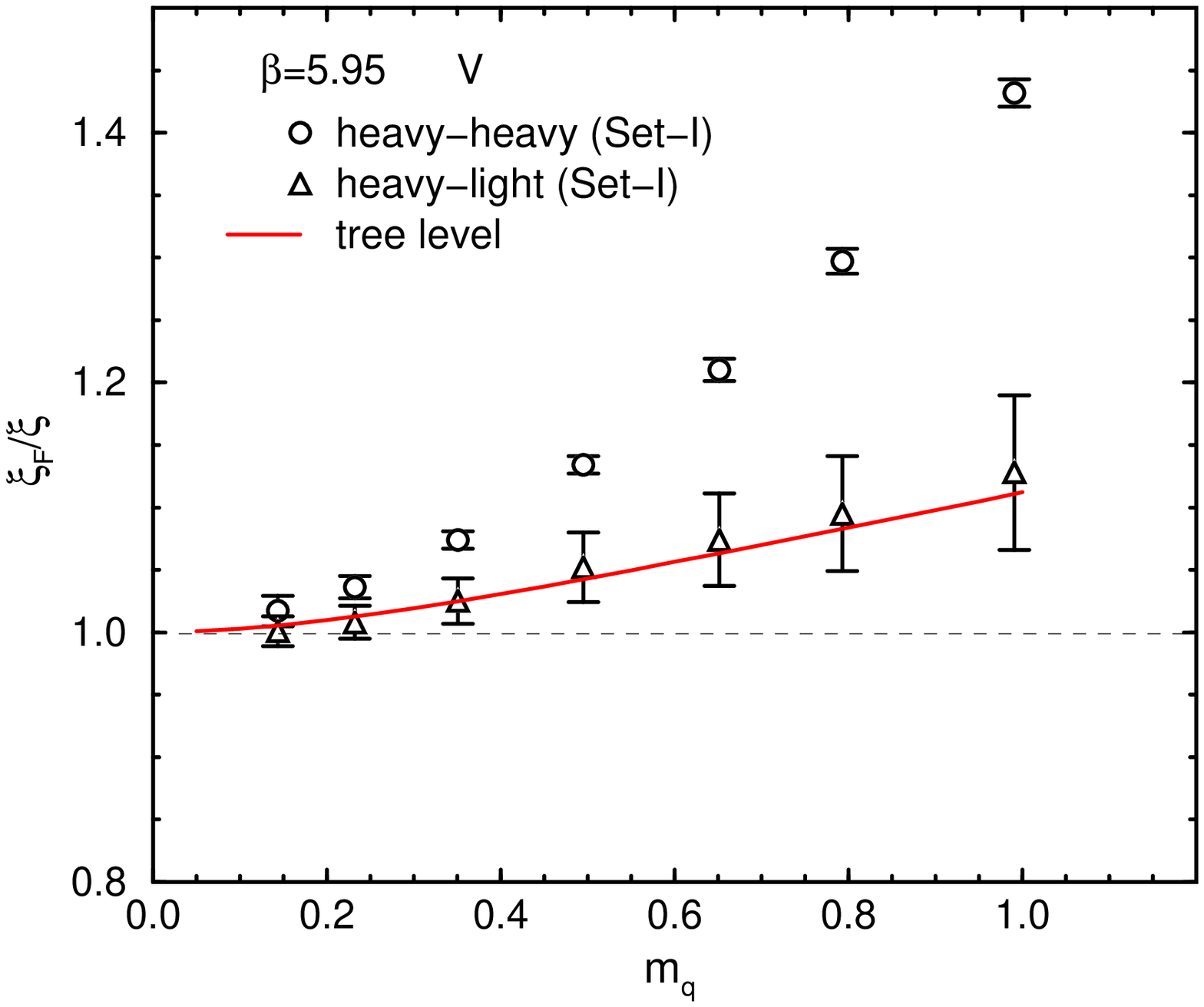}
\vspace{-0.6cm}
\caption{
Fermionic anisotropy determined nonperturbatively from the
dispersion relations of heavy-heavy and heavy-light
mesons.
The left and right panels show the results for
the pseudoscalar and vector channels, respectively.
The solid lines represent the tree level values according
to Eq.~(\ref{eq:xi_F_xi}).}
\label{fig:xi_F}
\end{figure*}

Figure~\ref{fig:xi_F} displays the heavy quark mass dependence of
$\xi_F$ for {\it Set-I}.
The horizontal axis is the bare quark mass $m_q$ in temporal
lattice units. 
The behaviors of $\xi_F$'s are well in accord with the expectation
in Sec.~\ref{sec:formulation}.
For quantitative discussion, let us consider the case that the
required accuracies to define the $a_\sigma^{hh\hf hl}$ and
$a_\sigma^{hl\hf ll}$ are 2 \%, namely
$\epsilon_{acc} = \epsilon_{acc} = 0.02$.
$\xi_F$'s from the heavy-heavy and heavy-light mesons
disagree beyond this accuracy at $m_q > 0.2$ ($a_{\sigma} m_Q > 0.8$).
Therefore one must keep $m_q < 0.2$ to avoid large systematic
uncertainty in the heavy quarkonia.
On the other hand, the $\xi_F$ from the heavy-light mesons 
are rather close to $\xi$, and $\gamma_F^*(m_q=0)$ can be applied
up to $m_q \simeq 0.3$ within presently required accuracy.
For quark mass larger than this value, the discrepancy between
$\xi_F$'s from the pseudoscalar and vector mesons gradually
grows beyond the statistical error.
This signals the growth of systematic error.
In the region of $m_q < 0.3$ such effect is sufficiently small.

For the charm quark mass, the present lattice spacing
$a_\sigma$ is already well less than $a_\sigma^{hl\hf ll}$
so that the $\gamma_F$ tuned for massless quark is applicable
to the charmed hadron systems.
On the other hand, bottom quark mass is not the case, and one
need finer lattice spacing or larger anisotropy $\xi$.
The present lattice would be also sufficient for the charmonium
states, since the region $m_q < 0.2$ also covers the charm quark mass.
Another striking feature is that the observed $\xi_F$'s for
heavy-light mesons are close to the tree level expectation.
This implies that the deviation of $\xi_F$ from $\xi$ may be
largely removed by the tree level tuning of $\gamma_F$.
Such an approach would work for the spectroscopy of hadrons 
containing single bottom quark.
This is a good alternative procedure to the mass dependent
calibration using heavy-light meson, since the statistical
error of $\xi_F$ from heavy-light mesons rapidly grows
as heavy quark mass.

Now we summarize the result for the {\it Set-II}.
The heavy-heavy meson satisfies the relativity relation
by definition.
However, the heavy-light meson dispersion relation
violates the relativity relation so that $\xi_F/\xi$ deviates from
unity toward smaller value as heavy quark mass increases.
This implies that the mass dependent tuning cannot absorb the
discrepancy between the $\xi_F$'s from the heavy-heavy and
heavy-light meson systems.
For the quark mass of $m_q < 0.2$, the result for {\it Set-II} are
consistent with the result for {\it Set-I}.
In that region the heavy-heavy meson can be successfully described,
and therefore the result of calibration with heavy-heavy mesons
is also valid.

We now observe the inconsistency among the binding energies
of heavy-heavy, heavy-light, and light-light mesons
discussed in Refs.~\cite{SAnomaly1,SAnomaly2}.
The inconsistency is probed by a quantity
\begin{equation}
I \equiv \frac{ 2 \delta M_{hl}
                 -( \delta M_{hh} + \delta M_{ll}) }{2M_{2hl}},
\end{equation}
where $\delta M = M_2 - M_1$, $M_1$ and $M_2$ are the rest and kinetic 
masses of the meson, respectively.
The subscripts $hh$, $hl$, and $ll$ represent the quark contents
of each meson ($h$ for heavy and $l$ light quarks).
We neglect the last term in the numerator, since the
calibration of light quark mass region requires that
$\delta M_{ll}$ vanishes.
Since the rest and kinetic quark masses cancel in each kind
of mass, nonvanishing $I$ represents the inconsistency
in the binding energy, namely the dynamical effect.
The anomalous behavior of $I$ in large kinetic mass region
was first reported in Ref.~\cite{SAnomaly1} for the
$O(a)$ improved quark action on isotropic lattice.
Ref.~\cite{SAnomaly2} explained that this behavior
originates from the $O((ap)^2)$ discretization effect in the heavy
quarkonium system and estimate the size of $I$ with the help
of potential model analysis.
It results in $I\simeq -0.5$ at $a M_{2hl}\simeq 3.2$, which
is in good agreement with the result in \cite{SAnomaly1}.

\begin{figure}
\includegraphics[width=8.8cm]{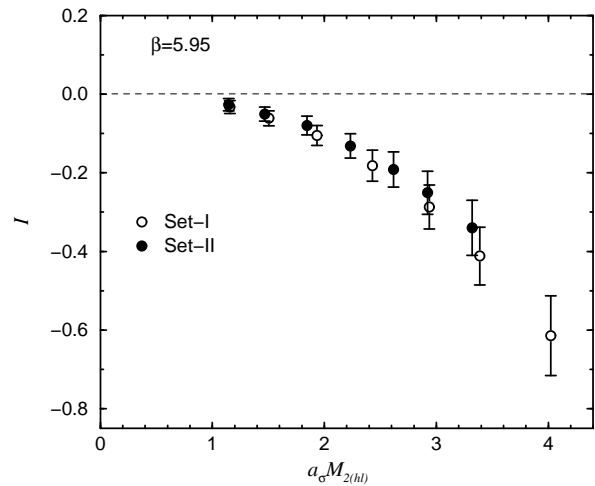}
\vspace{-0.8cm}
\caption{
The inconsistency among the binding energies of heavy-heavy,
heavy-light and light-light mesons.
The horizontal axis is in spatial lattice units.}
\label{fig:Sanomaly}
\end{figure}

Figure~\ref{fig:Sanomaly} displays the results of $I$
for {\it Set-I} and {\it II} for the pseudoscalar channel.
The behavior of $I$ is quite similar to that in
Refs.~\cite{SAnomaly1,SAnomaly2}, as is expected from that
it originates from $O((a_\sigma p)^2)$ error and therefore
is not improved by the anisotropy.
The behavior of $I$ of two sets is very similar to each other.
It means that the inconsistency cannot be eliminated by
tuning anisotropy parameters with either heavy-heavy or heavy-light
system, and a kind of universal quantity to signal the
failure in consistent description of heavy quarkonium systems.
The $I$ rapidly deviate from zero for the heavy quark mass larger than
our lightest one.
This is consistent with the above observation of $\xi_F$.

These results indicate that for $a_\tau m_Q > 0.2$ the present action
fails to describe the dynamics of heavy quarkonium, and cannot be
reinterpreted by the subtraction of quark mass effect.
Thus we need to either give up to apply the present framework 
for such a quark mass region, or to improve the action by incorporating
higher order correction terms.
The remaining part of this section therefore focus on the
heavy-light meson spectrum.

\subsection{Heavy-light meson spectrum}

Now we turn our attention to the heavy-light meson,
which is of our main interest.
According to the heavy quark expansion, the spin flipping interaction
of heavy quark in the heavy-light systems is of $O(1/m_Q)$,
and hence the hyperfine splitting of mesons is proportional to the
meson mass at the leading order.
In the heavy-light systems, the large mass of heavy quark is not
important for the dynamics.
Employing the Fermilab formulation, this character is taken into
account and the the correct heavy quark expansion is in terms
of kinetic mass \cite{EKM97}.
Therefore, the hyperfine splitting, which is measured as the
difference of rest masses, are expected to be reciprocally
proportional to the kinetic meson mass.
If this is the case, {\it Set-I} and {\it Set-II} should show the similar
behaviors, up to the $O(a^2)$  systematic uncertainty.

\begin{figure}
\includegraphics[width=8.8cm]{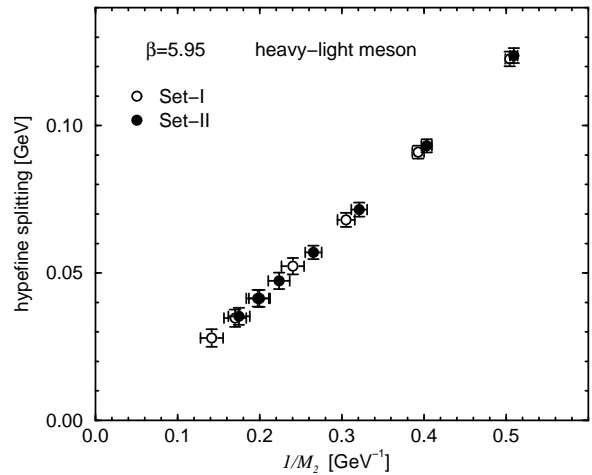}
\vspace{-0.8cm}
\caption{
The hyperfine splitting of heavy-light meson versus
the inverse of the heavy-light meson kinetic mass.
The physical scale is set by the hadronic radius $r_0$.
The error in $1/M_2$ is the systematic one estimated from
the discrepancy of $\xi_F$ from pseudoscalar and vector mesons.}
\label{fig:HFhl}
\end{figure}

The hyperfine splitting $m_V-m_{PS}$ is displayed
in Figure~\ref{fig:HFhl}.
The horizontal axis is spin averaged kinetic meson mass inverse
in physical units defined through the hadronic radius $r_0$.
Since the most serious uncertainty in the kinetic mass is
from the systematic uncertainty in $\xi_F$, we estimate
this error as
\begin{equation}
 \delta M_2 = 2 M_2 |\delta \xi_F| / \xi_F ,
\end{equation}
where $\delta \xi_F = \xi_F^{(V)} - \xi_F^{(PS)}$,
and hence it does not include the statistical errors of $\xi_F$ and
$M_2$.
On the other hand, the error associated to the hyperfine splitting
is the statistical one.
Although the data shows a linear dependence in the heavy quark mass
region, there appears small negative intercept.
We regard this small discrepancy with the heavy quark expansion
as the Lorentz symmetry breaking effect brought in
the meson dispersion relation and of $O(a^2)$.
For fixed physical quark mass, this effect is expected to disappear
linearly in $a^2$ toward the continuum limit.
The results of {\it Set-I} and {\it Set-II} clearly show similar
behavior, and therefore the above interpretation of Fermilab
formulation works well up to the relativity breaking effect
represented by $\delta \xi_F$.
For more quantitative analysis, one must quantify the size of
$O(a^2)$ systematic errors and observe how they disappear toward
the continuum limit.
This is out of the scope of this paper.

\section{Conclusion}
  \label{sec:conclusion}

In this paper, we investigated an applicability of the anisotropic
lattice quark action in heavy quark mass region, on the quenched
lattice with $a_\sigma^{-1}\simeq 1.6$ GeV and renormalized anisotropy
$\xi=4$.
The anisotropy is expected to extend the region in which the
parameters in the action tuned for massless quark are applicable
to precision computations of heavy-light matrix elements.
For this purpose, we measured the heavy-heavy and heavy-light meson
masses and dispersion relations to observe the relativity breaking
effect.
The calculation was carried out for two sets of parameters,
{\it Set-I} and {\it Set-II}.
{\it Set-I} adopts the values at the massless limit,
while in {\it Set-II} the bare anisotropy is tuned using the
heavy-heavy mesons.
Our main results are as follows.

(a) In the quark mass region $a_\tau m_Q < 0.2$,
the observed fermionic anisotropy $\xi_F$'s are consistent 
for heavy-heavy and heavy-light mesons as well as
for {\it Set-I} and {\it Set-II} parameters within 2 \% accuracy.
This implies that present framework is applicable to both these
systems even without tuning of anisotropy parameter.
Beyond this region, the action fails to describe the heavy quarkonium
state correctly, as expected from the tree level analysis of
the quark dispersion relation.

(b) The mass dependence of the renormalized anisotropy for the 
heavy-light mesons with  $\gamma_F^*$ at the massless limit ({\it Set-I}) 
is so small that one can exploit the massless tuning
in the region of $a_\tau m_Q < 0.3$ with less than 2\% errors.
This result is essentially important for our strategy, since it
implies that the parameters tuned at the massless limit is directly
available for this mass region, which already contains charm quark
mass with the present lattice and can be extended to the bottom quark
with development of computational resources in successive decade.

(c) For  $a_\tau m_Q > 0.3$, relativity breaking effect in the 
heavy-light mesons seems to grow as a function of the heavy quark 
mass. This is signaled by the discrepancy of
$\xi_F$'s from the pseudoscalar and vector mesons, although the
present statistics is not enough for a quantitative estimate of this effect.
In the scaling of hyperfine splitting, we also found small discrepancy
with the expectation from the heavy quark expansion, which is
considered as $O(a^2)$ systematic effect.
It is important to quantify these effects and to observe how they
disappear toward the continuum limit, for future high precision
computations  with this approach.

We therefore conclude that the anisotropic lattice quark action 
satisfies the required characters (i)--(iii) for high precision
calculations of heavy-light matrix elements mentioned in
Introduction.
Among above results, the small mass dependence of anisotropy
parameter is in particular quite encouraging for further
development of the framework in this direction.

One of the promising strategies is to calibrate the parameters
in the action at the massless limit and use them for all masses.
This should work perfectly for $a_{\tau} m_Q < 0.3$ which is already
sufficient to describe the $D$ meson systems. 
If one wants to treat $B$ meson systems, one has to work with heavier
quark, $a_{\tau} m_Q > 0.3$, in which case the mass dependent 
errors cannot be neglected.
However, since the mass dependence 
is small, it can be interpreted as an $O(a^2)$ error and can be
removed by taking the continuum limit.
Alternatively, one can also apply the genuine Fermilab approach
for the bottom quark.
In this case, the mass dependences of the renormalization coefficients
are the source of systematic errors.
Nevertheless, as long as one obtains such coefficients nonperturbatively
in the massless limit first, and use the one-loop perturbation theory
only to compute the mass dependent corrections, the perturbative
error can be much better controlled compared
to the Fermilab approach on the isotropic lattice.
The nice agreement between the observed $\xi_F$ and the tree level
expectation suggest that this idea is also promising in the
bottom quark mass region.

On the other hand, present level of systematic errors as well as
the statistical one is not sufficient for these ends.
As a most significant example, one need to eliminate the
$O(\alpha a)$ error using nonperturbative renormalization technique
in the massless limit \cite{Kla98a}, and confirm that this suffices
also for heavy quark mass region treated in this paper.

\section*{Acknowledgments}

We thank T.~Umeda for useful discussions.
The simulation was done on
NEC SX-5 at Research Center for Nuclear Physics, Osaka University and
Hitachi SR8000 at KEK (High Energy Accelerator Research Organization).
H.~M. is supported by Japan Society for the Promotion of Science
for Young Scientists.
T.~O. is supported by the Grant-in-Aid of the Ministry
of Education No. 12640279.
A.~S. is supported by the center-of-excellence (COE) program at
Research Center for Nuclear Physics, Osaka University.

\end{document}